\documentclass[superscriptaddress,amsmath,amssymb,aps]{revtex4-2} 
\usepackage{array}
\usepackage{upgreek}
\usepackage{natbib}
\usepackage{caption}
\usepackage[version=4]{mhchem}
\usepackage[pdftex]{graphicx}
\usepackage{graphicx}
\usepackage{dcolumn}
\usepackage{color}
\usepackage{bm}
\usepackage{hyperref}
\usepackage{siunitx}
\usepackage{gensymb}

\begin{document}

\preprint{APS/123-QED}

\title{Target-depth sensing with metasurface-encoder integrated optoelectronic neural network}

\author{Shuo Wang}
\author{Deyu Zhu}
\affiliation{Institute of Microelectronics, Chinese Academy of Sciences, Beitucheng west road 3, 100029, Beijing, China}
\affiliation{University of Chinese Academy of Sciences, Beijing 101408, China.}

\author{Chenjie Xiong}
\author{Bin Hu}
\affiliation{School of Optics and Photonics, Beijing Institute of Technology, Beijing 100081, China}
\affiliation{National Key Laboratory on Near-surface Detection, Beijing 100072, China.}

\author{Chunqi Jin}
\affiliation{State Key Laboratory of Integrated Optoelectronics, College of Electronic Science and Engineering, Jilin University, Changchun 130012, China}
\author{Yu Wang}
\author{Chengjun Zou}\email{Corresponding author: zouchengjun@ime.ac.cn}
\affiliation{Institute of Microelectronics, Chinese Academy of Sciences, Beitucheng west road 3, 100029, Beijing, China}

\date{\today}

\begin{abstract}
Accurate and real-time sensing of targets in three-dimensional (3D) environments is essential for modern machine vision, underpinning emerging technologies such as autonomous systems, robotic manipulation, augmented reality, and intelligent surveillance. However, state-of-the-art 3D sensing approaches typically rely on complex postprocessing of multi-view images or LiDAR point clouds, resulting in considerable computational load, power consumption, and latency.
To address these challenges, we propose a metasurface-encoder integrated optoelectronic neural network architecture that compresses 3D information into two-dimensional images by encoding depth using double-helix point spread function generated by a metasurface. The depth-encoded images are captured with a conventional monocular camera and subsequently processed by a lightweight shadow ResNet neural network.
We experimentally validate the proposed architecture on the MNIST and Vehicle-Image datasets, achieving high accuracy simultaneously in target classification and depth estimation, thereby enabling real-time target tracking. The framework is readily extendable to other depth- or angle-encoding metasurfaces for multidimensional compression and detection. Our results demonstrate the effectiveness of the meta-optic-encoder/electronic-decoder paradigm in significantly reducing network complexity and computational burden while maintaining strong performance for smart vision sensory applications.
\end{abstract}

\keywords{Optoelectronic neural network, depth estimation, object classification, metasurface encoders}

\maketitle

\section{Introduction}
Accurate perception of three-dimensional (3D) environments is fundamental to modern machine vision, enabling the reconstruction of 3D scenes that underpin a broad range of emerging technologies, including robotics \cite{yu2023brain}, autonomous navigation \cite{chen2023self}, augmented reality \cite{krichenbauer2017augmented,geroimenko2023augmented}, and intelligent surveillance \cite{yan2019introduction}. Over the past decades, substantial progress has been achieved in 3D sensing for depth recovery. Stereo vision estimates depth passively by triangulating corresponding features observed from multiple viewpoints \cite{saxena2007depth}, while structured light infers depth by analyzing the deformation of projected spatial patterns on a scene \cite{fanello2016hyperdepth}. Time-of-flight techniques provide pixel-level depth measurements by detecting the propagation delay or phase shift of reflected light, whereas light detection and ranging (LiDAR) determines distance by measuring the round-trip travel time or frequency shifts of scanned laser points \cite{liu2018tof}. These techniques varies in terms of their measurement accuracy, application scenarios, and postprocessing complexity.

Despite these advances, significant challenges persist as emerging technologies become increasingly interactive, integrated, and intelligent. For example, humanoid robots must perceive dynamic 3D scenes in real time while simultaneously performing dexterous tasks such as precision assembly. Autonomous vehicles require rapid reconstruction of their surrounding environments and corresponding control decisions within millisecond-scale latency. These demanding capabilities must be realized on resource-constrained platforms with continually tightening energy budgets. However, existing 3D sensing approaches typically rely on high-power active illumination and computationally intensive post-processing algorithms \cite{wang2021challenges}, leading to substantial latency, elevated power consumption, and increased system complexity. These limitations are further exacerbated by the rapid integration of artificial intelligence (AI), which drives explosive growth in sensory data volume and inference complexity \cite{wang2021multi}, while progress in Moore's law continues to slow \cite{lundstrom2022moore}. Consequently, there is an urgent need for 3D sensing technologies that provide not only accurate depth and object recognition, but also fast, low-latency, and energy-efficient perception in dynamic real-world environments.

In conventional processing architectures, sensors such as cameras primarily perform data acquisition, while nearly all subsequent information processing is delegated to electronic hardware. By contrast, integrating free-space optical encoding or computation directly during data acquisition can significantly reduce processing complexity, thereby lowering energy consumption and system latency \cite{choi2025free, froch2025computational}. This so-called ``pre-sensor'' processing paradigm \cite{huang2024pre} has recently attracted considerable attention. At the macroscopic scale, diffractive masks \cite{yan2019fourier,shi2022loen}, spatial light modulators (SLMs) and digital micromirror devices (DMDs) \cite{miscuglio2020massively,zhou2021large,yuan2023training} have been widely employed to encode or modulate incident light fields, serving as key components for optical convolutional processing. Diffractive multilayer devices have even been trained to emulate electronic deep neural networks, enabling functionalities such as image classification \cite{lin2018all}, noise-robust information encryption \cite{icsil2024all,zhang2024memory}, communication multiplexing \cite{fang2024orbital}, and spectrometers \cite{wang2024opto,cheong2024broadband}. At the microscopic scale, metasurfaces have been engineered to perform a variety of specialized image-processing operations, including differentiation \cite{zhou2020flat,wang2023metalens,zhang2025momentum}, correlation \cite{wang2022single}, and phase-contrast imaging \cite{ji2022quantitative}. Moreover, metasurface-based convolutional kernels \cite{zheng2022meta,zheng2024multichannel} can be directly integrated with camera photodetectors to enable fast and energy-efficient convolution operations \cite{luo2024meta}. Despite this significant progress, most existing efforts have focused on two-dimensional (2D) image processing, and the extension of pre-sensor optical computing concepts to 3D vision sensing \cite{shi2021multiple,yan2024nanowatt} remains relatively underexplored.

To address this gap, we propose and experimentally demonstrate a metasurface-encoder-integrated optoelectronic neural network (\textbf{MONN}) architecture capable of simultaneously performing target classification and depth (distance from the metasurface) estimation with low latency. The metasurface encoder compresses 3D volumetric information into 2D images by encoding depth with a designed double-helix point spread function (DH-PSF) \cite{jin2019dielectric,shen2023monocular}. The encoded images are captured by a conventional camera and subsequently processed by a pre-trained shadow ResNet neural network \cite{wen2020transfer}. We evaluate the proposed system using the MNIST handwritten digit dataset and a Vehicle-Image dataset (see \textbf{Experimental Section}), achieving >97.6\% accuracy in target classification and depth estimation errors around 1\% within the designed measurement range, comparable to sophisticated AI models. Real-time tracking of the target trajectories is also experimentally demonstrated. The robustness of the approach is further validated by changing image size and partially occluding the targets, while maintaining high detection accuracy. Although the demonstrated metasurface encoder represents a specific encoding strategy, the proposed processing architecture is readily extendable to a broad range of alternative optical encoding designs. Overall, this work establishes a scalable optoelectronic paradigm for low-latency, energy-efficient 3D perception by jointly leveraging metasurface-based optical encoding and alleviating the complexity of deep neural networks, offering a promising pathway toward next-generation intelligent vision systems.

\section{Results and Discussions}
\subsection{Metasurface encoder design and numerical validation of the MONN architecture}
The MONN processing architecture is depicted in \textbf{Figure~\ref{F1}a}. The target image and its depth information are optically encoded by the specially designed metasurface before captured by the CMOS camera. A narrow band-pass filter is placed before the camera for selecting the designed working wavelength of 785~nm. After optical to electronic conversion, the encoded image data is fed into the ResNet neural network for further processing, which simultaneously functions as the target classifier and the corresponding depth decoder. 

Firstly, we present the design of the depth-encoding metasurface and its experimental characterization. Many different encoding principles have been proposed for depth detection. For in-house experimental validation, we design a metalens-integrated DH-PSF phase mask \cite{jin2019metasurface} as the depth encoder (see Section~S1.1 of the Supporting Information (SI) for design details). Under plane-wave illumination, this metasurface generates two focal spots whose relative orientation rotates monotonically along the propagation axis, enabling quasi-linear depth encoding. With the distance between the phase mask and the CMOS camera fixed at 115~mm, the designed depth measurement range spans from 430~mm to 930~mm. The DH-PSF metasurface is realized using silicon nanoposts as the fundamental meta-units to faithfully implement the calculated phase profile. Figures~\ref{F1}b and \ref{F1}c show the corresponding unit-cell parameters and the photo together with a scanning electron micrograph (SEM) of the fabricated metasurface, which has an overall size of 4.8$\times$4.8~mm$^2$ for each array. The fabrication was outsourced to the commercial foundry Tianjin H-Chip Technology Group using electron beam lithography (EBL) and four metasurfaces were fabricated with different EBL dose variations (see \textbf{Experimental Section} for detailed fabrication procedures). Figure~\ref{F1}d presents the experimentally characterized depth-encoding performance, revealing a monotonic relationship between the double-helix rotation angle and the light-source (target) distance (See Section~S1.2 in the SI for specific experimental setup). The measured PSF intensity distributions for selected depth positions (labeled as 1, 11, 21,...) are summarized in Figure~\ref{F1}e. Additionally, measured images of the digit ``6'' with and without the depth encoding is also shown in Figure~\ref{F1}e, where two overlapping replicas of the digit clearly correspond to its distance-dependent encoding by the metasurface.
\begin{figure}[t!]
\centering
  \includegraphics[width=1\linewidth]{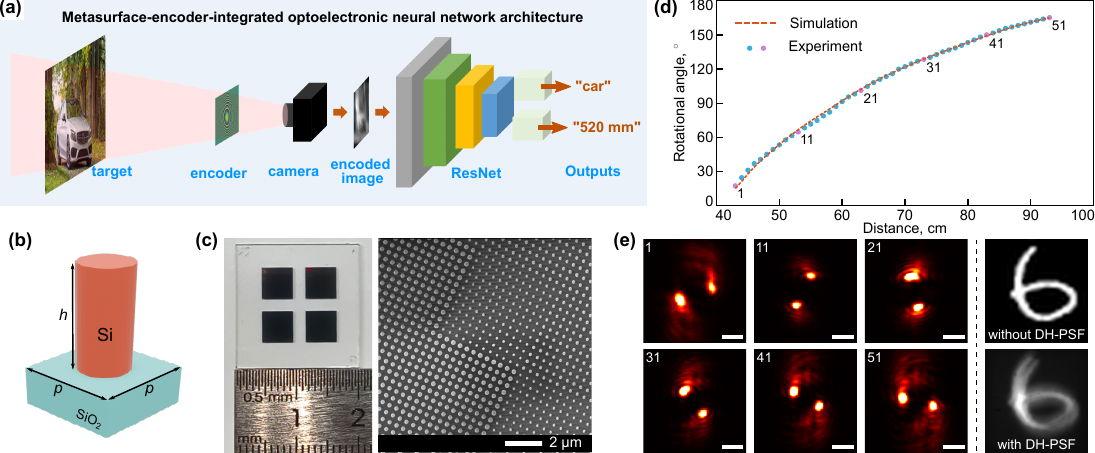}%
  \caption{Concept of the MONN processing architecture and the metasurface encoder results. (a) Illustration of the MONN processing architecture showing the whole processing flow. (b) Unit cell of the metasurface. Here the period $p=400$~nm and nanopost height $h=550$~nm; the diameter of the silicon nanopost varies from 80~nm to 205~nm covering 2$\pi$ phase modulation. (c) Photo of the fabricated DH-PSF metasurface and corresponding SEM image. (d) Simulated (i.e. designed) and measured depth-encoding against the double-helix rotation angle with respect to the horizontal plane. In total, 51 depth positions were measured. (e) The measured PSF intensities correspond to the labeled positions (purple dots) in (d) are presented. The white scale bars mark a length of 100~$\upmu$m. On the right side, the measured image for a digit ``6'' with and without the depth-encoding are presented.}
  \label{F1}
\end{figure}

Next, based on the experimentally measured metasurface response, we numerically construct and evaluate the performance of the target-depth sensing MONN architecture. The MNIST handwritten digit dataset and the Vehicle-Image dataset are employed for validation. Here we note that: explicit 3D object models are not used for training or testing the MONN, primarily due to the limited open availability of large-scale 3D datasets and, more critically, the practical challenges associated with fabricating a substantial number of physical 3D objects for experimental verification. Instead, we generate the training and testing data by placing 2D images from above-mentioned datasets at varying depth positions. The key distinction between using 2D and 3D targets for validation lies in the preservation of local depth information: while true 3D objects inherently has local depth variation (i.e. 3D shapes etc.), which is neglected for 2D images. However, for estimating the general distance from an object to the observation point, the local depth variations are often ignored in many scenarios.
\begin{figure}[t!]
\centering
  \includegraphics[width=1\linewidth]{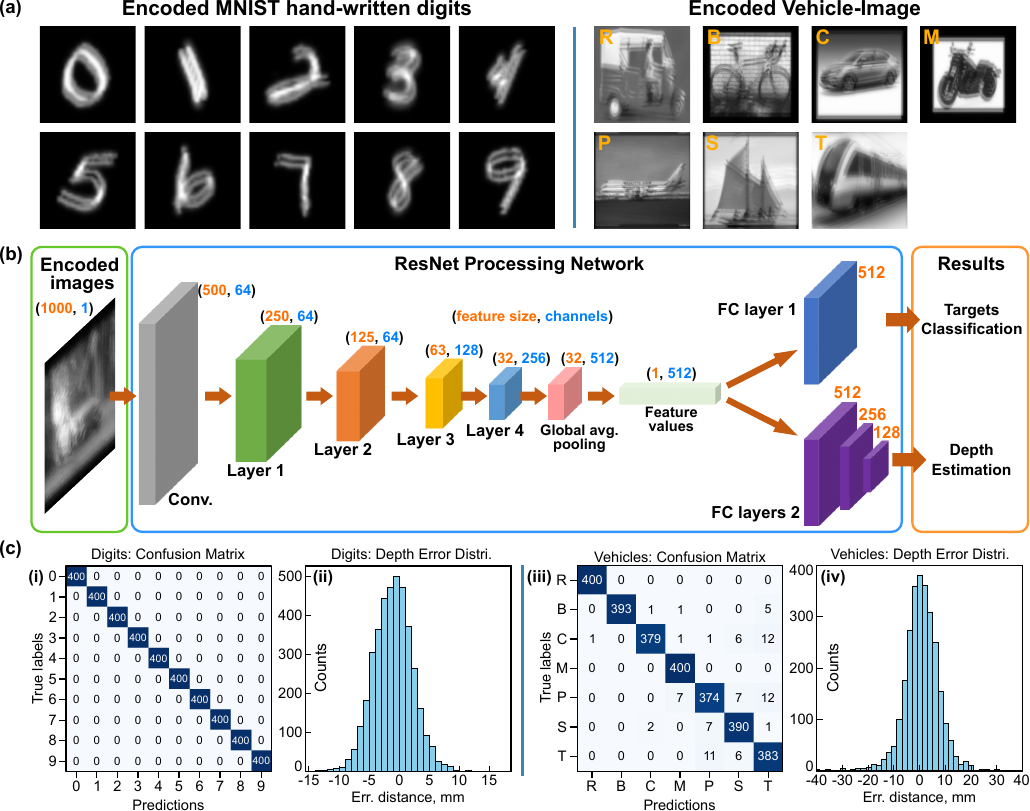}%
  \caption{Numerical validation of the MONN depth-object sensing architecture. (a) Preparation of training datasets by convoluting the MNIST and Vehicle images with measured DH-PSF profiles. (b) Illustration of the adapted ResNet processing network. (c) Simulation evaluation results of the trained model on target classification (i,iii) and depth estimation error distribution (ii,iv).}
  \label{F2}
\end{figure}

To generate the training datasets for MONN, the original images were first convolved with the measured DH-PSFs to encode depth information and produce the corresponding training samples. In total, 40,000 digit images and 28,000 vehicle images were selected as the base datasets. Each image was then convolved with 20 randomly selected (nearly uniformly distributed) DH-PSFs corresponding to the spatial sampling points shown in Figure~\ref{F1}d, yielding a large and evenly distributed training dataset. Representative encoded inputs are shown in \textbf{Figure~\ref{F2}a}, clearly exhibiting the characteristic overlapped double-helix patterns. Because the DH-PSF metasurface includes a metalens phase component, a thin-lens imaging effect naturally occurs, in which the apparent target size decreases with increasing object distance from the metasurface. To ensure that the MONN learns depth information solely from the DH-PSF encoding, we randomly rescale all convolved images to remove image-size cues, effectively suppressing the lens-based depth signal that conventional depth-from-defocus networks often rely on.

Figure~\ref{F2}b illustrates the MONN architecture adapted from ResNet-34. The encoded grayscale inputs have a spatial resolution of 1000$\times$1000 pixels. After feature extraction through the ResNet backbone, the network produces a $1\times1\times512$ feature representation, which is passed to two separate fully connected heads: one for target classification and the other for continuous depth regression. The network was trained using the AdamW optimizer. Cross-entropy loss was applied to the classification head, and root-mean-square error (RMSE) loss to the depth-regression head. The total loss was defined as a weighted sum of the two, with a weight ratio of 1:10 (classification:regression) to compensate for the slower convergence of the regression branch. Training was performed on an NVIDIA GeForce RTX 4090 GPU for nearly 200 epochs, after which the model achieved strong performance on both tasks.

For evaluation, 4,000 MNIST images and 2,800 vehicle images that were not used in training were selected to construct the testing datasets. Each image was again convolved with 20 randomly chosen DH-PSFs to match the training format. The testing results are summarized in Figure~\ref{F2}c. For the MNIST dataset, the network achieved a classification accuracy of 100\% (Figure~\ref{F2}c-i). Depth-estimation errors were predominantly within 10~mm (Figure~\ref{F2}c-ii), with a mean absolute error (MAE) of 2.7 mm and an RMSE of 3.4 mm, demonstrating highly accurate depth sensing from encoded single images. For the vehicle dataset, the classification accuracy reached 97.11\% (Figure~\ref{F2}c-iii). The depth-estimation errors were mostly within 20 mm, with an MAE of 5.5 mm and an RMSE of 8.3 mm. Importantly, the depth estimation error distributions are nearly uniform across the entire designed depth measurement range of 430–930 mm, indicating a robust performance. Overall, the normalized RMSE values with respect to the full measurement range are 0.68\% for the MNIST dataset and 1.66\% for the vehicle dataset, confirming that the DH-PSF encoded MONN achieves highly accurate, stable, and scalable 3D sensing performance for different targets and depth conditions.

\subsection{Experimental evaluation on target-depth sensing}
After numerically implementing and examining the MONN processing architecture, we further conducted experimental validation. \textbf{Figure~\ref{F3}a} illustrates the customized measurement setup, consisting of a LED light source (center wavelength $\lambda=785$~nm), a movable target holder, the metasurface encoder, and a camera (CS165MU from Thorlabs) equipped with a 10~nm narrow bandpass filter (FBH05785-10 from Thorlabs). 
\begin{figure}[t!]
\centering
  \includegraphics[width=0.9\linewidth]{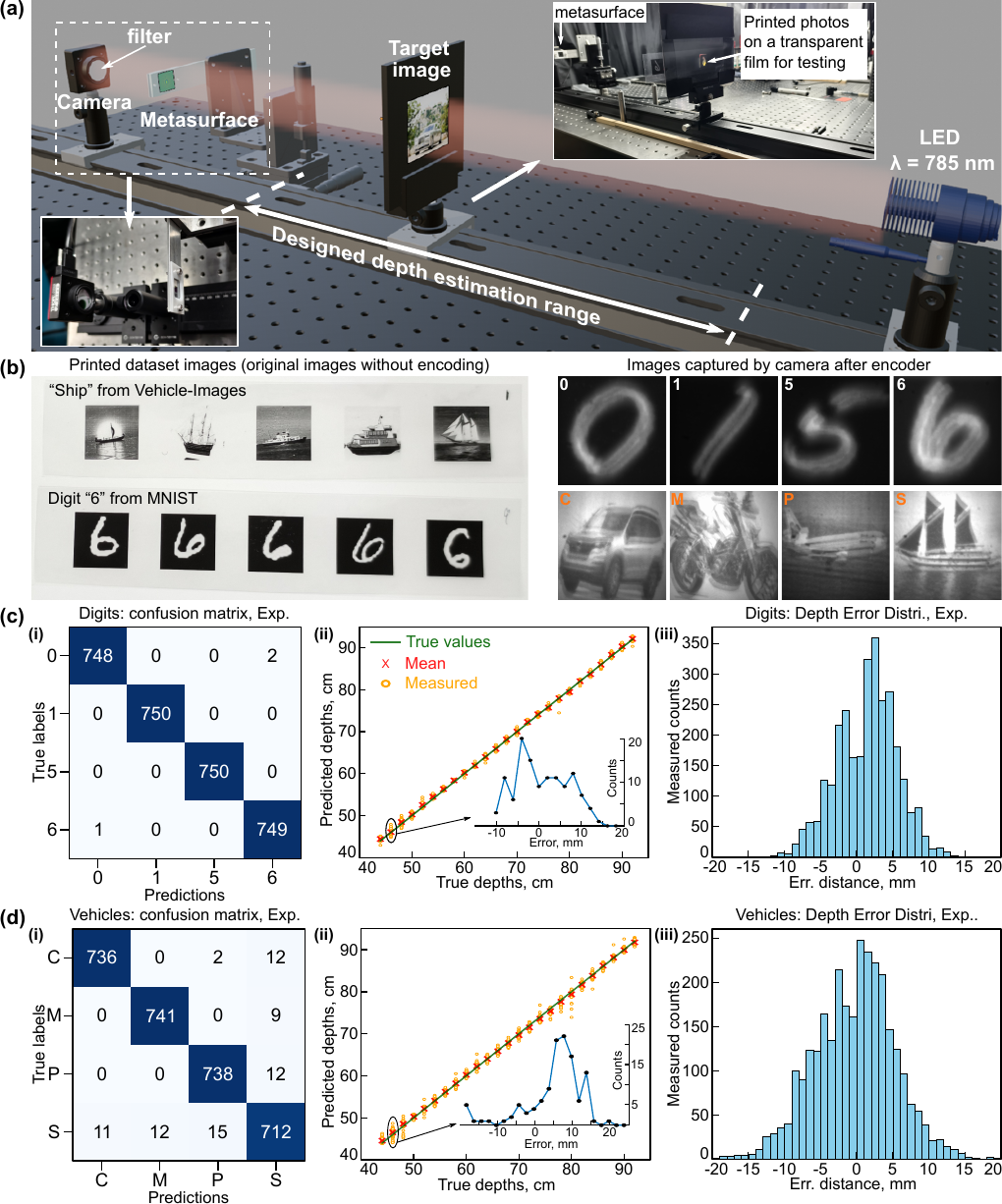}%
  \caption{Experimental examination of the MONN processing architecture. (a) Illustration of the experimental setup for depth-object sensing. (b) Examples of printed images for MONN evaluation (left) and recorded images by camera (right). (c,d) Experimental evaluation results for (c) MNIST digit targets and (d) Vehicle-Image targets including: (i) confusion matrix showing classification accuracy; (ii) Tested depth estimation at different depth positions; (iii) Error distribution for target depth estimation.}
  \label{F3}
\end{figure}
The two insets present photographs of the metasurface encoder mounted in front of the camera and the movable target holder on the rail. To accurately reproduce the input targets at different depths, images from the two datasets were printed on transparent films (hereafter referred to as ``photos'' to distinguish them from the digital dataset images) and mounted onto the target holder for LED illumination. For simplicity, and to avoid excessive printing, four image categories from each dataset were selected. Within these selected categories, we randomly chose 150 digit images and 180 vehicle images for printing. Some examples of printed photos and their corresponding experimentally recorded encoded images are shown in Figure~\ref{F3}b.

During experimental validation, we first tested the sensing performance using the previously trained network based on numerically convolved images (see Figure~\ref{F2}a). However, its classification accuracy and depth estimation performance were extremely poor (see Section~S2 in the SI). This degradation is mainly attributed to substantial differences between the numerically generated training images and real recorded data, particularly in terms of background noise and illumination uniformity. For example, the LED illumination introduces an approximately Gaussian intensity profile across the target plane, resulting in noticeable brightness gradients. Furthermore, regions corresponding to pure black pixels in the dataset images appear non-zero in the captured photos due to nonzero transmission, random scattering, and camera noises. These discrepancies hinder the ability of the network trained solely on synthetic data to generalize to real measurements. To address this challenge, we constructed a new training dataset using experimentally recorded encoded photos. Specifically, 120 MNIST digit photos and 150 vehicle photos were placed at 20 randomly selected (quasi-uniformly distributed) depth positions within the designed sensing range, and their encoded measurements were recorded to build training datasets. Using these datasets, we retrained the MONN following the network architecture shown in Figure~\ref{F2}b. The remaining 30 printed photos from each dataset were reserved for performance evaluation.

For testing, all evaluation photos were positioned at 25 depths between 44~cm and 92~cm, with a step size of 2~cm, and tested for both category classification and depth estimation. The results are summarized in Figure~\ref{F3}c and \ref{F3}d. In target classification, extremely high accuracies of 99.9\% for handwritten digits and 97.6\% for vehicles were achieved, as shown in Figure~\ref{F3}c-i and \ref{F3}d-i. For depth retrieval, the estimated depths show excellent consistency with the ground truth (Figure~\ref{F3}c-ii, \ref{F3}d-ii). At each measurement position, 120 distinct encoded photos (four categories, 30 images per category) were evaluated. The MNIST photos achieved MAE and RMSE values of 3.68~mm and 4.53~mm, respectively, while the corresponding errors for vehicles were 4.60~mm and 6.13~mm. In the corresponding figures, two insets are included to show the error distribution at one measurement position. Such distributions are distorted from Gaussian distribution to certain degrees. However, the overall depth error statistics, plotted in Figure~\ref{F3}c-iii and \ref{F3}d-iii, follow near-Gaussian distributions, with the majority of measurements falling within 10~mm of error. These results clearly demonstrate that the MONN architecture maintains robust sensing performance under realistic optical conditions, highlighting its potential for practical 3D perception applications.

\subsection{Real-time targets' depth sensing}
\begin{figure}[t!]
\centering
  \includegraphics[width=0.85\linewidth]{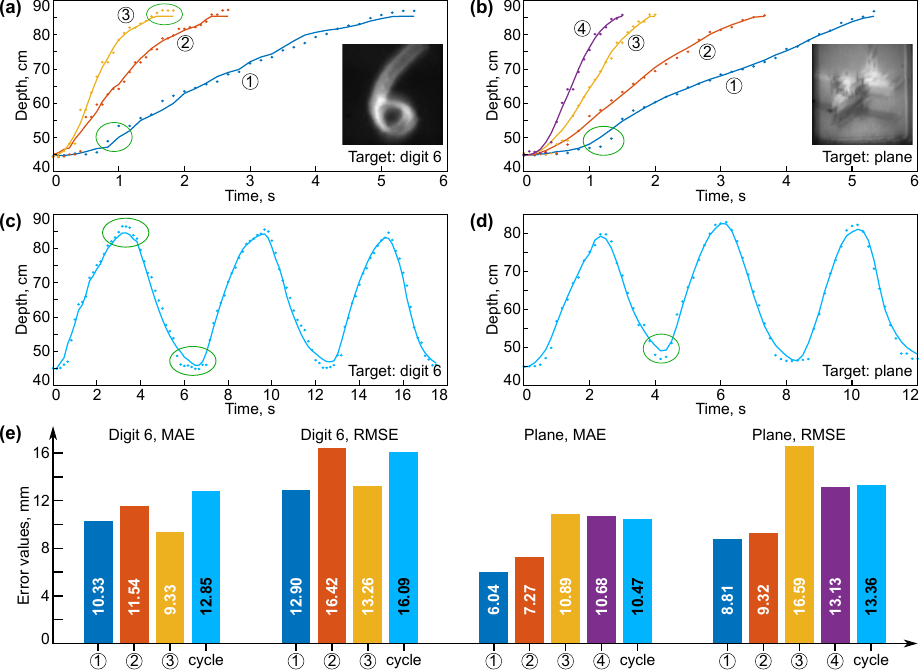}%
  \caption{Real-time depth sensing based on the MONN architecture. Solid lines indicate the reference target positions recorded by an iPhone, while dots represent the depths estimated by the MONN. (a,b) Tracking results for (a) the digit ``6'' and (b) the ``plane'' target moving at different velocities. The inset figures show the encoded target images captured by the camera. (c,d) Tracking results for cyclic back-and-forth motions of (c) the digit ``6'' and (d) the ``plane''. (e) MAE and RMSE results of the real-time depth measurements.}
  \label{F4}
\end{figure}
To evaluate the real-time sensing capability of the MONN architecture, we further conducted a real-time target tracking experiment using the trained networks and the same characterization setup shown in Figure~\ref{F3}a. The processing accuracy and latency were systematically analyzed, and in particular the real-time depth sensing is the focus here. Two randomly selected test targets, a photo of the digit ``6'' and a photo of a ``plane'', were used in the experiment. The targets were mounted on a wheeled platform and manually moved at different speeds (See Section~S3 in the SI), while both the ground-truth, time-dependent depth positions and the MONN-estimated real-time positions were recorded. During the measurements, the camera integration time was set to 10~ms, corresponding to a theoretical maximum frame rate of 100 frames per second (FPS). The reference depth positions were captured using an iPhone~14 mounted on the wheeled platform (see Figure~S4a in the SI), operating at 60~FPS. Meanwhile, the real-time outputs of the MONN were recorded using screen-capture software at a maximum frame rate of 30~FPS. Videos documenting the manual motion of the targets were also recorded. All video materials are compiled in the accompanying experiments recording movies (see \textbf{Supporting Information Section}).

\textbf{Figure~\ref{F4}} summarizes the measured results corresponding to the recorded videos. The two targets were manually translated from approximately 45~cm to 85~cm at different speeds, and the measured depth trajectories are compared in Figure~\ref{F4}a and \ref{F4}b. In these plots, the solid curves represent the reference depths recorded by the iPhone, while the discrete markers (dots) denote the MONN-estimated results. To improve plot clarity, depth values extracted from the MONN screen recordings were sampled every four to five frames. In addition, cyclic back-and-forth motion of both targets was tested, with the corresponding results shown in Figure~\ref{F4}c and \ref{F4}d. Overall, the MONN-estimated depths show excellent agreement with the reference measurements, particularly when the targets move at approximately constant velocities. Larger discrepancies are mainly observed during periods of target acceleration, as highlighted by the green-circled regions. These deviations arise from the combined effects of network processing latency (discussed in the next paragraph) and target acceleration, which amplify temporal mismatches between the estimated and reference positions. Figure~\ref{F4}e further summarizes the MAEs and RMSEs of the network estimations relative to the reference measurements. In general, these error metrics almost doubled than those obtained under static conditions. Intuitively, the estimation error should increase as the target move at higher velocities. However, no clear or systematic dependence of the error magnitude on the target moving speeds is observed across the tested range.

Although the camera integration time was set to only 10~ms, the experimentally measured average processing cycle time, defined as the interval between two successive MONN outputs, was approximately 100~ms. The trained MONN model was deployed on a Dell XPS~15 9530 laptop equipped with an NVIDIA 4050D GPU and 6~GB memory. The inference time of the neural network for a single 1000$\times$1000-pixel image is approximately 30~ms (from PyCharm timing). The additional $\sim$60~ms latency mainly arises from data readout from the camera to the laptop and then the image resizing operations, which are performed on the CPU. Such processing overheads are typical for laptop-based implementations and can be substantially reduced through hardware acceleration strategies, such as in-sensor processing or electronic neuromorphic computing architectures \cite{chen2023all}.
Here, we note that a direct performance comparison with purely electronic deep-learning models for simultaneous target classification and absolute depth estimation was not implemented. Since these models rely on substantially deeper and more complex network architectures as well as require more training data \cite{cs2018depthnet,bochkovskii2024depth,wen2025foundationstereo}, they are unable to achieve comparable processing speed and accuracy based on the same hardware platform. By contrast, the key contribution of the MONN framework lies in its ability to replace computationally intensive electronic models with a compact and lightweight neural network, enabling highly accurate absolute depth estimation (with $\sim$1\% error) and reliable target classification at significantly reduced computational cost.

\subsection{Evaluations on MONN robustness}
We further evaluate the robustness of the MONN processing architecture by examining its performance under variations in target size and partial target occlusion. For these tests, only the Vehicle-Image photos are used, since handwritten digits are particularly sensitive to occlusion and thus unsuitable for a systematic robustness analysis.
\begin{figure}[b!]
\centering
  \includegraphics[width=0.9\linewidth]{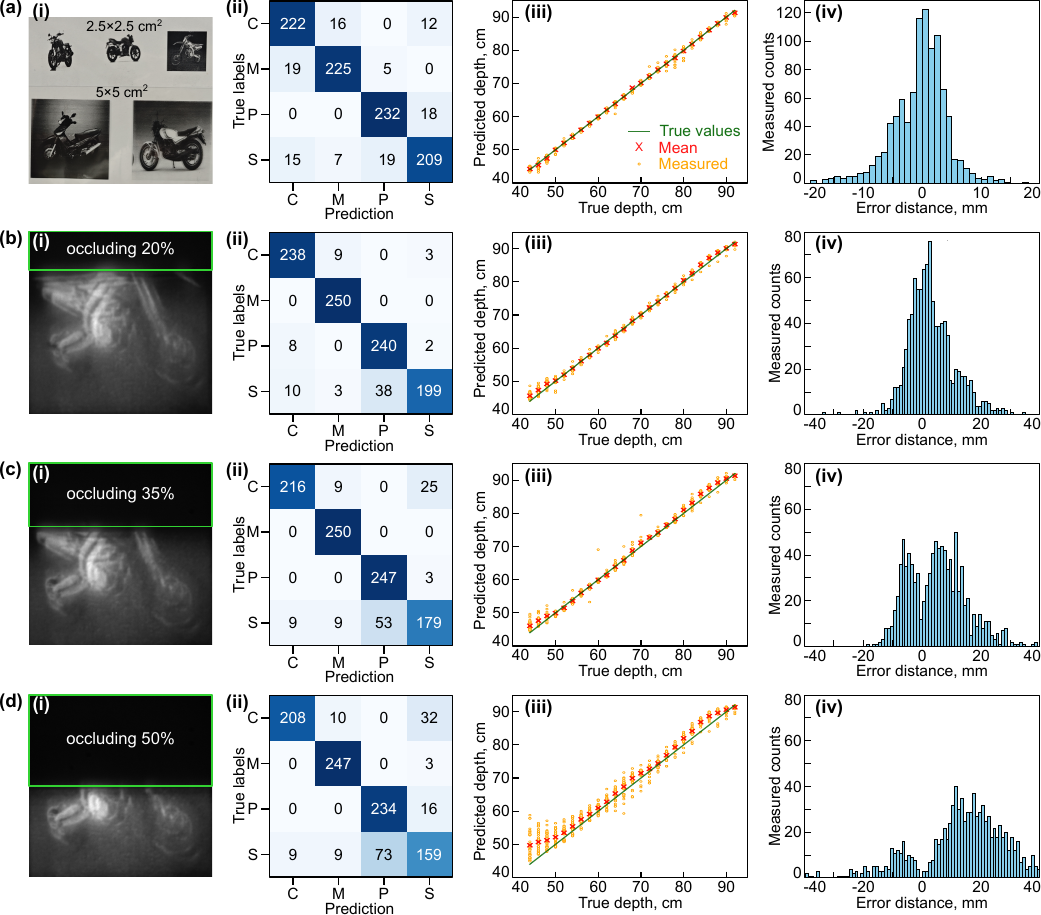}%
  \caption{Robustness analysis of the MONN processing architecture. (a) Performance stability under target size variation. (i) Examples of enlarged target photos compared with the previously used smaller ones. (ii) Confusion matrix for target classification. (iii) Corresponding depth estimation results and (iv) the associated depth error distributions.
 (b–d) MONN estimation results under partial target occlusion from the top direction: (b) 20\%, (c) 35\%, and (d) 50\% occlusion. For each case, sub-figures (ii)–(iv) present the (ii) target classification confusion matrices, (iii) depth estimation results, and (iv) corresponding depth error distributions.}
  \label{F5}
\end{figure}

To assess the influence of target size, we aim to exclude possible lens-based zooming effects in depth estimation and to evaluate the stability of the system under scale variations. Specifically, we print larger target photos with a size of 5$\times$5~mm$^2$, as shown in \textbf{Figure~\ref{F5}a-i}, and test them using the same experimental configuration illustrated in Figure~\ref{F3}a. For evaluation, ten enlarged photos from each category (car, motor, plane, and ship) are prepared and again positioned at 25 depth locations ranging from 44~cm to 92~cm, with a step size of 2~cm. The corresponding statistical results are summarized in Figure~\ref{F5}a-ii to a-iv. In terms of target classification, the accuracy for the enlarged photos decreases to 88.89\%, which is 8.7\% lower than that obtained with the smaller-size targets. In contrast, the depth-sensing performance remains largely comparable: the resulting MAE and RMSE are 4.35~mm and 11.57~mm, respectively, as shown in Figure~\ref{F5}a-iii and a-iv. Similar to the results in Figure~\ref{F3}d-ii, relatively high  fluctuations are observed around a depth of approximately 80~cm, although the overall fluctuations are more pronounced for the larger targets. These results indicate that while classification accuracy is moderately affected by target scaling, the depth estimation capability of the MONN remains relatively robust against size variations, possibly due to the optical encoding of the depth information.

Next, we investigate the impact of partial occlusion on system performance. As illustrated in Figure~\ref{F5}b–d, the targets are occluded from the top by approximately 20\%, 35\%, and 50\%, respectively. For category classification, high accuracies are still achieved under moderate occlusion, although the number of misclassifications increases with occlusion severity. The corresponding classification accuracies are 92.7\%, 89.2\%, and 84.8\%, as shown in Figure~\ref{F5}b-ii to d-ii. For depth estimation, a clear degradation in accuracy is observed as the occlusion level increases, as evidenced in from Figure~\ref{F5}b-iii,iv to \ref{F5}d-iii,iv. The error distributions broaden progressively with increasing occlusion. Specifically, the (MAE, RMSE) values are (6.20~mm, 8.54~mm), (9.25~mm, 12.38~mm), and (21.31~mm, 27.32~mm) for 20\%, 35\%, and 50\% occlusion, respectively. While the performance remains acceptable for occlusion levels up to 35\%, more severe occlusion leads to a substantial degradation in depth estimation accuracy, rendering results beyond 50\% occlusion unreliable. This degradation is expected and could potentially be mitigated by incorporating more diverse occlusion scenarios into the training dataset. Although the current optical encoding may have provided partial robustness against occlusion, a deeper understanding of how optical encoding strategies can be optimized to preserve depth information under occluded conditions requires further investigations.

\section{Conclusion}
In conclusion, we have demonstrated a MONN processing framework for real-time 3D sensing that combines optical encoding with lightweight neural processing to achieve simultaneous target classification and absolute depth estimation based on monocular camera observation. Experimental results show high classification accuracy, and depth-sensing precision up to around 1\% over the designed operation range, as well as stable real-time tracking of moving targets.
Robustness tests further indicate that the MONN maintains reliable depth estimation under moderate target size variations and partial occlusions, highlighting the importance of optical encoding design in preserving depth information. 
Different from purely electronic deep-learning approaches that typically rely on substantially more complex network architectures, the proposed MONN demonstrates how physics-based optical preprocessing can significantly reduce computational burden while maintaining high accuracy by reducing the network complexity. These results underscore the potential of metasurface-assisted neural sensing as a promising pathway toward low-latency, energy-efficient 3D perception, with future improvements expected from optimized optical encoding strategies and hardware-efficient implementations.

\section*{Experimental Section}
\textbf{Fabrication of DH-PSF metasurface:}
The designed metasurface layout was outsourced to the commercial foundry Tianjin H-Chip Technology Group for fabrication. Commercially available JGS-1 quartz glass chip (20$\times$20~mm$^2$) was selected as the substrate. First, a 550~nm thick amorphous silicon was deposited by plasma-enhanced chemical vapor deposition (PECVD). Then, the chip was coated by a chromium layer of 30~nm thickness by ion-beam deposition (KURT J.LESKER Labline PVD 75), and subsequently a 150~nm thick layer of electron-beam resist (ZEP520) was applied. After exposure with electron-beam lithography system (JEOL JBX-9500 FS), the resist was developed. The chromium layer was then etched with ion-beam etching (OXFORD PlasmaPro 100 Cobra180). The defined pattern of the chromium mask was transferred to the silicon layer by reactive ion etching (OXFORD PlasmaPro 100 Cobra180). A very shallow over-etch into the silica substrate occurred during the etching process. Finally, the residual resist and the chromium mask were dissolved in acetone and chromium etching solution (ceric ammonium nitrate + perchloric acid), respectively.

\textbf{Preparation of datasets:}
The MNIST handwritten digit dataset and the {\color{blue}\href{https://www.kaggle.com/datasets/mohamedmaher5/vehicle-classification}{Vehicle-Image dataset}} were selected for training and testing. To ensure compatibility with the experimental characterization setup, selected images were printed onto A4-sized transparent plastic sheets and cut into long strips; such transparent sheets are commonly used for overhead projectors. Each standard printed image has a size of 25$\times$25~mm$^2$, while enlarged versions used in specific robustness tests have dimensions of 50$\times$50~mm$^2$. For the digit dataset, 150 images were randomly selected from the MNIST classes 0, 1, 5, and 6 and printed. For the vehicle dataset, 180 images were randomly selected from the categories car, motor, ship, and plane. Among these, 120 digit photos and 150 vehicle photos were placed at 20 randomly selected positions along the translation rail (Fig.~\ref{F3}a) to record the encoded images and construct the training dataset. The remaining 30 printed photos from each dataset were reserved for network evaluation.

\section*{Supporting Information}
Supporting Information includes a document and two videos. The document summarizes: 1) Design and characterization of the metasurface encoder. 2) Performance of MONN trained by numerically generated data. 3) Real-time targets' depth sensing setup. The videos record the real-time target-depth sensing of a digit photo (6) and a vehicle photo (plane).

\begin{acknowledgments}
C.Z. acknowledges the funding support from the CAS Pioneer Talents Program. This work was also supported by the National Nature Science Foundation of China (No. 62305372).
\end{acknowledgments}


\newpage
\section*{Supporting Information}

\renewcommand{\thefigure}{S\arabic{figure}}
\renewcommand{\thetable}{S\arabic{table}}
\renewcommand{\theequation}{S\arabic{equation}}
\renewcommand{\thepage}{S\arabic{page}}
\setcounter{figure}{0}

\section*{S1. Design and characterization of the metasurface encoder}
\subsection*{S1.1 Design of the metalens-integrated DH-PSF metasurface}
The metasurface for depth encoding employs the strategy of double-helix point spread function (DH-PSF) creating two rotatable focal spots when illuminated by a plane wave incidence. To ensure high-efficiency operation, the metasurface is made by high-transmission silicon nano-post meta-units, with the design process focusing primarily on optimizing the phase distribution. We follow the design methodology outlined in~\cite{jin2019metasurface}, where the initial structure is composed of five Gaussian-Laguerre (GL) modes with indices (1,1), (3,5), (5,9), (7,13), and (9,17), each with equal magnitude. To incorporate the function of imaging, a lens phase profile (numerical aperture being 0.055) is superimposed on the initial DH-PSF phase mask, with the resulting phase distribution depicted in Figure~\ref{F1}a.
\begin{figure}[b!]
\centering
  \includegraphics[width=1\linewidth]{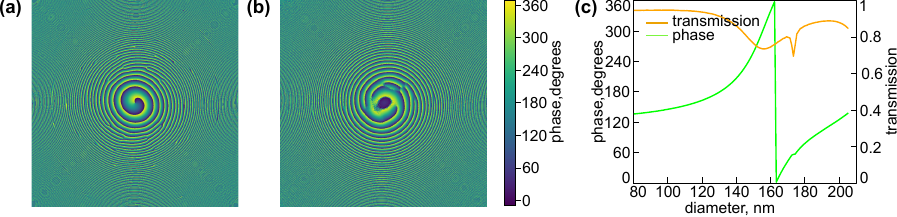}%
  \caption{Designed of the DH-PSF metasurface encoder. (a) Initial designed phase distribution of the DH-PSF metasurface. (b) Optimized phase distribution of the DH-PSF metasurface. (c) Simulated transmission phase and magnitude of silicon nano-posts.}
  \label{F1}
\end{figure}

Next, the initial phase distribution is optimized with an iterative method close to the Gerchberg–Saxton (GS) algorithm. The optimization criterion is to achieve a 180\degree~rotation of the DH-PSF at nine distinct positions across the designed depth encoding range. The optimized phase distribution is shown in Figure~\ref{F1}b. To achieve this, the objective function of the iterative method is defined as the sum of the light intensity distributions at the nine positions. Each intensity distribution is modeled by two center-symmetric Gaussian main lobes, as defined in the following equation:
\begin{align}
O_{\text{Gauss}} &= \exp\left\{ - \left[ \frac{(x - x_a)^2}{\omega_{G1}^2} + \frac{(y - y_a)^2}{\omega_{G2}^2} \right] \right\}
    + \exp\left\{ - \left[ \frac{(x - x_b)^2}{\omega_{G1}^2} + \frac{(y - y_b)^2}{\omega_{G2}^2} \right] \right\} \tag{S1} \\
O_{\text{step}} &=
 \begin{cases}
    1, & (|x-x_a|\leq \omega_1 \cap |y-y_a|\leq \omega_2) \cup (|x-x_b|\leq \omega_1 \cap |y-y_b|\leq \omega_2)\\
     \gamma, & \text{otherwise}
 \end{cases} \tag{S2}
\end{align}
Here, \(\text{O}_{\text{Gauss}}\) refers to the double-Gaussian distribution, and \(\text{O}_{\text{step}}\) represents the step function. The parameters \(\omega_{G1,G2}\) denote the widths of the Gaussian beam waist along the long and short axes of the main lobes, while \(s_{1,2}\) represent the widths of the step function. $(x_a, y_a)$ and $(x_b, y_b)$ correspond to the locations of the maximum values of the main lobes, respectively. The attenuation factor $\gamma$, which is less than 1, describes the degree of attenuation.

Finally, we convert the metasurface phase distribution into the corresponding metasurface structure. As previously described, we use silicon nano-posts as the fundamental meta-units to accurately implement the calculated phase profile at 785~nm. The intensity and phase distribution of the meta-units are shown in Figure~\ref{F1}c. The diameter of the meta-units varies from 80~nm to 205~nm, covering a $2\pi$ phase modulation range while exhibiting high transmission efficiency.

\subsection*{S1.2 Experimental characterization of the DH-PSF metasurface}
To validate the design PSF of the metasurface, we characterize its PSF with the setup shown in Figure~\ref{F2}, which includes a LED light source, a pinhole (300~$\upmu$m), the metasurface, and a camera. The whole setup is similar to that in Figure~3a. The distance between the metasurface and the camera is fixed at 115~mm. By gradually increasing the distance between the metasurface and the pinhole, the rotational DH-PSF responses are recorded, as illustrated in Figure~1d.
\begin{figure}[h!]
\centering
  \includegraphics[width=1\linewidth]{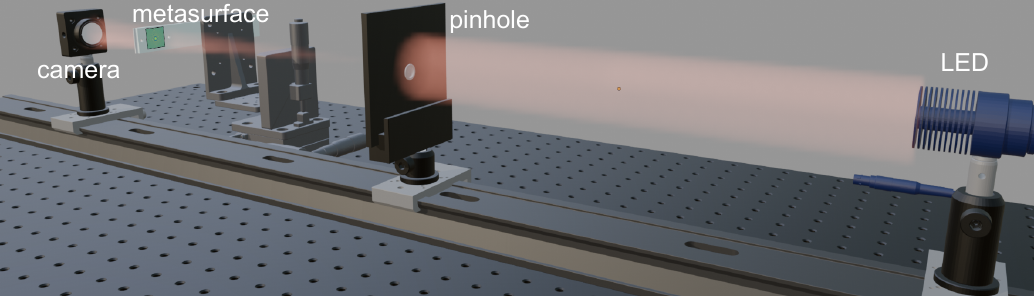}%
  \caption{Characterization of the DH-PSF metasurface.}
  \label{F2}
\end{figure}

\section*{S2. Performance of MONN trained by numerically generated data}
\begin{figure}[b!]
\centering
  \includegraphics[width=0.9\linewidth]{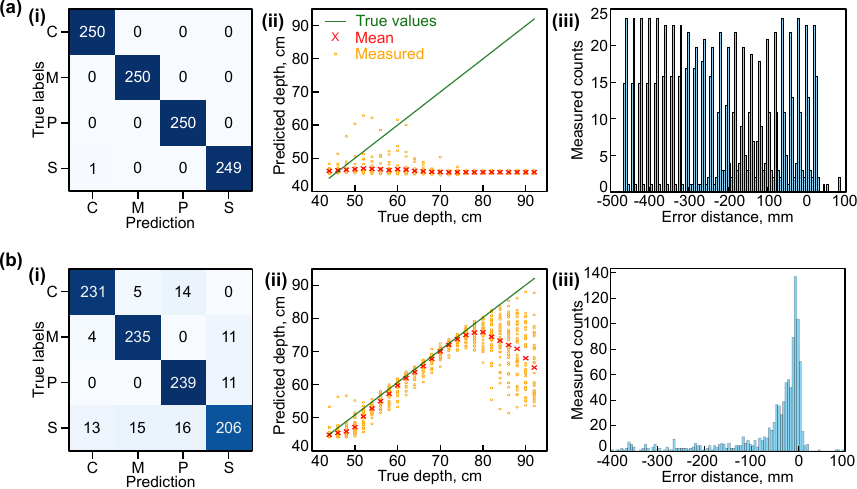}%
  \caption{Experimental examination of the MONN model trained by using the numerically generated images for (a) the digit dataset and (b) the vehicle image dataset. Results including (i) confusion matrix showing classification accuracy, (ii) tested depth estimation at different depth positions, and (iii) error distribution for target depth estimation.}
  \label{F3}
\end{figure}
The measurement results obtained using the MONN model trained on numerically generated datasets, constructed by convolving selected dataset images with the measured DH-PSF profiles, are presented here. All measurements were performed using the same experimental setup shown in Figure~\ref{F3}a. Figure~\ref{F3} summarizes the experimental results for the (a) MNIST digit dataset and the (b) Vehicle-Image dataset. As shown in Figure~\ref{F3}a-i and \ref{F3}b-i, high target classification accuracies are achieved for both datasets, reaching 99.9\% for digits and 91.1\% for vehicles.

In contrast, the depth-sensing performance differs markedly between the two datasets. For the digit dataset, depth estimation essentially fails, yielding nearly constant predicted depths around 45~cm regardless of the true target position (Figure~\ref{F3}a-ii). For the vehicle dataset, however, the model provides reasonably accurate depth estimates up to approximately 74~cm, as shown in Figure~\ref{F3}b-ii. Within this range (44–74~cm), the mean absolute error (MAE) and root-mean-square error (RMSE) are 16.9~mm and 24.0~mm, respectively. Beyond 74~cm, the deviation from the true depth increases rapidly.

We attribute this behavior to the effective zooming induced by the embedded lens phase of the metasurface, as discussed in the main text. When targets are closer to the system, the camera recorded images show the target sizes and light intensities more closely resemble those in the numerically generated training data. As the target distance increases, the projected image becomes smaller, leading to a stronger contribution from background LED illumination on the camera, which acts as noise for the trained network. This effect is also pronounced for the digit dataset, where the digit occupies only a small fraction of the image area at all depths, resulting in dominant background noise across the entire sensing range. In contrast, the MONN trained on experimentally recorded images implicitly learns such noise characteristics, enabling more robust depth estimation under realistic measurement conditions.

\section*{S3. Real-time targets' depth sensing setup}
Figure~\ref{F4} presents the experimental setup for the real-time target tracking measurements. Both the target photos and the iPhone are mounted on a wheeled platform (Figure~\ref{F4}a). The platform movement trajectories are defined by the wood rails shown in the figure. 

We also note that two rails were connected to make a whole railed optical measurement setup. The iphone recorded video clips show the targets move start from 1~cm. However, before the marked ``1~cm'', a offset distance of 44~cm need to be added.
\begin{figure}[h!]
\centering
  \includegraphics[width=0.9\linewidth]{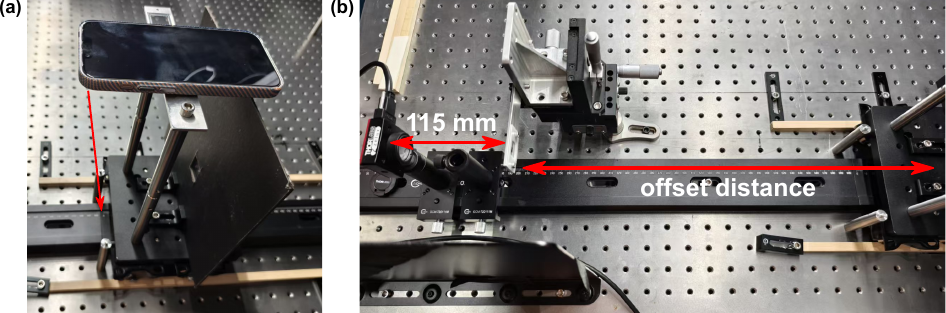}%
  \caption{Illustration of the real-time experimental scene. (a) A mounted iPhone to record reference depth while moving targets. (b) Offset distance different from the tick marks on the rail, which is used to obtain the real depth.}
  \label{F4}
\end{figure}


\begin{thebibliography}{10}
\providecommand{\url}[1]{\texttt{#1}}
\providecommand{\urlprefix}{URL }

\bibitem{yu2023brain}
F.~Yu, Y.~Wu, S.~Ma, M.~Xu, H.~Li, H.~Qu, C.~Song, T.~Wang, R.~Zhao, L.~Shi,
\newblock \emph{Science Robotics} \textbf{2023}, \emph{8}, 78 eabm6996.

\bibitem{chen2023self}
N.~Chen, F.~Kong, W.~Xu, Y.~Cai, H.~Li, D.~He, Y.~Qin, F.~Zhang,
\newblock \emph{Science Robotics} \textbf{2023}, \emph{8}, 76 eade4538.

\bibitem{krichenbauer2017augmented}
M.~Krichenbauer, G.~Yamamoto, T.~Taketom, C.~Sandor, H.~Kato,
\newblock \emph{IEEE transactions on visualization and computer graphics} \textbf{2017}, \emph{24}, 2 1038.

\bibitem{geroimenko2023augmented}
V.~Geroimenko,
\newblock \emph{Augmented reality and artificial intelligence: The fusion of advanced technologies},
\newblock Springer Nature, \textbf{2023}.

\bibitem{yan2019introduction}
W.~Q. Yan,
\newblock \emph{Introduction to intelligent surveillance: surveillance data capture, transmission, and analytics},
\newblock Springer, \textbf{2019}.

\bibitem{saxena2007depth}
A.~Saxena, J.~Schulte, A.~Y. Ng, et~al.,
\newblock In \emph{IJCAI}, volume~7. \textbf{2007} 2197--2203.

\bibitem{fanello2016hyperdepth}
S.~R. Fanello, C.~Rhemann, V.~Tankovich, A.~Kowdle, S.~O. Escolano, D.~Kim, S.~Izadi,
\newblock In \emph{Proceedings of the IEEE conference on computer vision and pattern recognition}. \textbf{2016} 5441--5450.

\bibitem{liu2018tof}
J.~Liu, Q.~Sun, Z.~Fan, Y.~Jia,
\newblock In \emph{2018 IEEE 3rd Optoelectronics Global Conference (OGC)}. IEEE, \textbf{2018} 185--190.

\bibitem{wang2021challenges}
Z.~Wang, M.~Menenti,
\newblock \emph{Frontiers in Remote Sensing} \textbf{2021}, \emph{2} 641723.

\bibitem{wang2021multi}
X.~Wang, C.~Wang, B.~Liu, X.~Zhou, L.~Zhang, J.~Zheng, X.~Bai,
\newblock \emph{Displays} \textbf{2021}, \emph{70} 102102.

\bibitem{lundstrom2022moore}
M.~S. Lundstrom, M.~A. Alam,
\newblock \emph{Science} \textbf{2022}, \emph{378}, 6621 722.

\bibitem{choi2025free}
M.~Choi, A.~Majumdar,
\newblock \emph{npj Nanophotonics} \textbf{2025}, \emph{2}, 1 36.

\bibitem{froch2025computational}
J.~E. Fr{\"o}ch, S.~Colburn, D.~J. Brady, F.~Heide, A.~Veeraraghavan, A.~Majumdar,
\newblock \emph{Optica} \textbf{2025}, \emph{12}, 6 774.

\bibitem{huang2024pre}
Z.~Huang, W.~Shi, S.~Wu, Y.~Wang, S.~Yang, H.~Chen,
\newblock \emph{Science Advances} \textbf{2024}, \emph{10}, 30 eado8516.

\bibitem{yan2019fourier}
T.~Yan, J.~Wu, T.~Zhou, H.~Xie, F.~Xu, J.~Fan, L.~Fang, X.~Lin, Q.~Dai,
\newblock \emph{Physical Review Letters} \textbf{2019}, \emph{123}, 2 023901.

\bibitem{shi2022loen}
W.~Shi, Z.~Huang, H.~Huang, C.~Hu, M.~Chen, S.~Yang, H.~Chen,
\newblock \emph{Light: Science \& Applications} \textbf{2022}, \emph{11}, 1 121.

\bibitem{miscuglio2020massively}
M.~Miscuglio, Z.~Hu, S.~Li, J.~K. George, R.~Capanna, H.~Dalir, P.~M. Bardet, P.~Gupta, V.~J. Sorger,
\newblock \emph{Optica} \textbf{2020}, \emph{7}, 12 1812.

\bibitem{zhou2021large}
T.~Zhou, X.~Lin, J.~Wu, Y.~Chen, H.~Xie, Y.~Li, J.~Fan, H.~Wu, L.~Fang, Q.~Dai,
\newblock \emph{Nature Photonics} \textbf{2021}, \emph{15}, 5 367.

\bibitem{yuan2023training}
X.~Yuan, Y.~Wang, Z.~Xu, T.~Zhou, L.~Fang,
\newblock \emph{Nature Communications} \textbf{2023}, \emph{14}, 1 7110.

\bibitem{lin2018all}
X.~Lin, Y.~Rivenson, N.~T. Yardimci, M.~Veli, Y.~Luo, M.~Jarrahi, A.~Ozcan,
\newblock \emph{Science} \textbf{2018}, \emph{361}, 6406 1004.

\bibitem{icsil2024all}
{\c{C}}.~I{\c{s}}{\i}l, T.~Gan, F.~O. Ardic, K.~Mentesoglu, J.~Digani, H.~Karaca, H.~Chen, J.~Li, D.~Mengu, M.~Jarrahi, et~al.,
\newblock \emph{Light: Science \& Applications} \textbf{2024}, \emph{13}, 1 43.

\bibitem{zhang2024memory}
Y.~Zhang, Q.~Zhang, H.~Yu, Y.~Zhang, H.~Luan, M.~Gu,
\newblock \emph{Science Advances} \textbf{2024}, \emph{10}, 24 eadn2205.

\bibitem{fang2024orbital}
X.~Fang, X.~Hu, B.~Li, H.~Su, K.~Cheng, H.~Luan, M.~Gu,
\newblock \emph{Light: Science \& Applications} \textbf{2024}, \emph{13}, 1 49.

\bibitem{wang2024opto}
Z.~Wang, H.~Chen, J.~Li, T.~Xu, Z.~Zhao, Z.~Duan, S.~Gao, X.~Lin,
\newblock \emph{Nanophotonics} \textbf{2024}, \emph{13}, 20 3883.

\bibitem{cheong2024broadband}
Y.~Z. Cheong, L.~Thekkekara, M.~Bhaskaran, B.~del Rosal, S.~Sriram,
\newblock \emph{Advanced Photonics Research} \textbf{2024}, \emph{5}, 6 2300310.

\bibitem{zhou2020flat}
Y.~Zhou, H.~Zheng, I.~I. Kravchenko, J.~Valentine,
\newblock \emph{Nat. Photon.} \textbf{2020}, \emph{14}, 5 316.

\bibitem{wang2023metalens}
S.~Wang, L.~Li, S.~Wen, R.~Liang, Y.~Liu, F.~Zhao, Y.~Yang,
\newblock \emph{Nano Lett.} \textbf{2023}, \emph{24}, 1 356.

\bibitem{zhang2025momentum}
K.~Zhang, S.~Wang, j.~Qiu, M.~Yang, T.~Liu, S.~Xiao, I.~Staude, T.~Pertsch, Y.~Wang, C.~Zou,
\newblock \emph{Adv. Opt. Mater.} \textbf{2025}, \emph{13}, 18 2500352.

\bibitem{wang2022single}
Z.~Wang, G.~Hu, X.~Wang, X.~Ding, K.~Zhang, H.~Li, S.~N. Burokur, Q.~Wu, J.~Liu, J.~Tan, C.~Qiu,
\newblock \emph{Nat. Commun.} \textbf{2022}, \emph{13}, 1 2188.

\bibitem{ji2022quantitative}
A.~Ji, J.-H. Song, Q.~Li, F.~Xu, C.-T. Tsai, R.~C. Tiberio, B.~Cui, P.~Lalanne, P.~G. Kik, D.~A. Miller, M.~L. Brongersma,
\newblock \emph{Nat. Commun.} \textbf{2022}, \emph{13}, 1 7848.

\bibitem{zheng2022meta}
H.~Zheng, Q.~Liu, Y.~Zhou, I.~I. Kravchenko, Y.~Huo, J.~Valentine,
\newblock \emph{Science Advances} \textbf{2022}, \emph{8}, 30 eabo6410.

\bibitem{zheng2024multichannel}
H.~Zheng, Q.~Liu, I.~I. Kravchenko, X.~Zhang, Y.~Huo, J.~G. Valentine,
\newblock \emph{Nature Nanotechnology} \textbf{2024}, \emph{19}, 4 471.

\bibitem{luo2024meta}
M.~Luo, T.~Xu, S.~Xiao, H.~K. Tsang, C.~Shu, C.~Huang,
\newblock \emph{Laser \& Photonics Reviews} \textbf{2024}, \emph{18}, 11 2300984.

\bibitem{shi2021multiple}
J.~Shi, L.~Zhou, T.~Liu, C.~Hu, K.~Liu, J.~Luo, H.~Wang, C.~Xie, X.~Zhang,
\newblock \emph{Optics Letters} \textbf{2021}, \emph{46}, 14 3388.

\bibitem{yan2024nanowatt}
T.~Yan, T.~Zhou, Y.~Guo, Y.~Zhao, G.~Shao, J.~Wu, R.~Huang, Q.~Dai, L.~Fang,
\newblock \emph{Science Advances} \textbf{2024}, \emph{10}, 27 eadn2031.

\bibitem{jin2019dielectric}
C.~Jin, M.~Afsharnia, R.~Berlich, S.~Fasold, C.~Zou, D.~Arslan, I.~Staude, T.~Pertsch, F.~Setzpfandt,
\newblock \emph{Advanced Photonics} \textbf{2019}, \emph{1}, 3 036001.

\bibitem{shen2023monocular}
Z.~Shen, F.~Zhao, C.~Jin, S.~Wang, L.~Cao, Y.~Yang,
\newblock \emph{Nature Communications} \textbf{2023}, \emph{14}, 1 1035.

\bibitem{wen2020transfer}
L.~Wen, X.~Li, L.~Gao,
\newblock \emph{Neural Computing and Applications} \textbf{2020}, \emph{32}, 10 6111.

\bibitem{jin2019metasurface}
C.~Jin, J.~Zhang, C.~Guo,
\newblock \emph{Nanophotonics} \textbf{2019}, \emph{8}, 3 451.

\bibitem{chen2023all}
Y.~Chen, M.~Nazhamaiti, H.~Xu, Y.~Meng, T.~Zhou, G.~Li, J.~Fan, Q.~Wei, J.~Wu, F.~Qiao, L.~Fang, Q.~Dai,
\newblock \emph{Nature} \textbf{2023}, \emph{623}, 7985 48.

\bibitem{cs2018depthnet}
A.~CS~Kumar, S.~M. Bhandarkar, M.~Prasad,
\newblock In \emph{Proceedings of the IEEE Conference on Computer Vision and Pattern Recognition Workshops}. \textbf{2018} 283--291.

\bibitem{bochkovskii2024depth}
A.~Bochkovskii, A.~Delaunoy, H.~Germain, M.~Santos, Y.~Zhou, S.~R. Richter, V.~Koltun,
\newblock \emph{arXiv preprint arXiv:2410.02073} \textbf{2024}.

\bibitem{wen2025foundationstereo}
B.~Wen, M.~Trepte, J.~Aribido, J.~Kautz, O.~Gallo, S.~Birchfield,
\newblock In \emph{Proceedings of the Computer Vision and Pattern Recognition Conference}. \textbf{2025} 5249--5260.
\end{thebibliography}
\end{document}